\documentclass[11pt]{amsart}
\usepackage{graphicx}
\usepackage{amscd}
\usepackage{amsbsy}
\usepackage{amsmath}
\usepackage{amstext}
\usepackage{amsfonts}
\usepackage{amssymb}
\newtheorem{theorem}{Theorem}
\theoremstyle{plain}

\newtheorem{corollary}{Corollary}

\newtheorem{definition}{Definition}

\newtheorem{lemma}{Lemma}

\newtheorem{proposition}{Proposition}
\newtheorem{remark}{Remark}

\numberwithin{equation}{section}

\makeindex

\begin{document}
\title[A Lecture on Shor's Factoring Algorithm]{A\\Lecture\\on\\Shor's Quantum Factoring Algorithm\\Version 1.1}
\author{Samuel J. Lomonaco, Jr.}
\address{Dept. of Comp. Sci. \& Elect. Engr.\\
University of Maryland Baltimore County\\
1000 Hilltop Circle\\
Baltimore, MD 21250}
\email{E-Mail: Lomonaco@UMBC.EDU}
\urladdr{WebPage: http://www.csee.umbc.edu/\symbol{126}lomonaco}
\thanks{This work was partially supported by ARO Grant \#P-38804-PH-QC and the L-O-O-P
Fund. The author gratefully acknowledges the hospitality of the University of
Cambridge Isaac Newton Institute for Mathematical Sciences, Cambridge,
England, where some of this work was completed. \ \ I would also like to thank
the other AMS\ Short Course lecturers, Howard Brandt, Dan Gottesman, Lou
Kauffman, Alexei Kitaev, Peter Shor, Umesh Vazirani and the many Short Course
participants for their support. \ (Copyright 2000.) \ }
\date{September 20, 2000}
\maketitle

\begin{abstract}
This paper is a written version of a one hour lecture given on Peter Shor's
quantum factoring algorithm. \ It is based on \cite{Ekert1}, \cite{Hoyer1},
\cite{Jozsa2}, \cite{Kitaev1}, and \cite{Shor1} .
\end{abstract}\tableofcontents

\section{Preamble to Shor's algorithm}

\quad\bigskip

There are cryptographic systems (such as RSA\footnote{RSA is a public key
cryptographic system invented by Rivest, Shamir, Adleman. \ Hence the name.
\ For more information, please refer to \cite{Stinson1}.}) that are
extensively used today (e.g., in the banking industry) which are based on the
following questionable assumption, i.e., conjecture: \ 

\bigskip

\noindent\textbf{Conjecture(Assumption). \ }\textit{Integer factoring is
computationally much harder than integer multiplication. \ In other words,
while there are obviously many polynomial time algorithms for integer
multiplication, there are no polynomial time algorithms for integer factoring.
\ I.e., integer factoring computationally requires super-polynomial time.}

\bigskip

This assumption is based on the fact that, in spite of the intensive efforts
over many centuries of the best minds to find a polynomial time factoring
algorithm, no one has succeeded so far. \ As of this writing, the most
asymptotically efficient \emph{classical} algorithm is the number theoretic
sieve \cite{Lenstra1}, \cite{Lenstra2}, which factors an integer $N$ in time
$O\left(  \exp\left[  \left(  \lg N\right)  ^{1/3}\left(  \lg\lg N\right)
^{2/3}\right]  \right)  $. \ Thus, this is a super-polynomial time algorithm
in the number $O\left(  \lg N\right)  $ of digits in $N$. \ 

\vspace{0.5in}

However, ... Peter Shor suddenly changed the rules of the game. \ \bigskip

Hidden in the above conjecture is the unstated, but implicitly understood,
assumption that all algorithms run on computers based on the principles of
classical mechanics, i.e., on \textbf{classical computers}. \ But what if a
computer could be built that is based not only on classical mechanics, but on
quantum mechanics as well? \ I.e., what if we could build a \textbf{quantum
computer}?

\bigskip

Shor, starting from the works of Benioff, Bennett, Deutsch , Feynman, Simon,
and others, created an algorithm to be run on a quantum computer, i.e., a
\textbf{quantum algorithm}, that factors integers in polynomial time! \ Shor's
algorithm takes asymptotically $O\left(  \left(  \lg N\right)  ^{2}\left(
\lg\lg N\right)  \left(  \lg\lg\lg N\right)  \right)  $ steps on a quantum
computer, which is polynomial time in the number of digits $O\left(  \lg
N\right)  $ of $N$. \ 

\vspace{0.5in}

\section{Number theoretic preliminaries}

\quad\bigskip

Since the time of Euclid, it has been known that every positive integer $N$
can be uniquely (up to order) factored into the product of primes. \ Moreover,
it is a computationally easy (polynomial time) task to determine whether or
not $N$ is a prime or composite number. \ For the primality testing algorithm
of Miller-Rabin\cite{Miller1} makes such a determination at the cost of
$O\left(  s\lg N\right)  $ arithmetic operations [$O\left(  s\lg^{3}N\right)
$ bit operations] \ with probability of error $Prob_{Error}\leq2^{-s}$. \ 

\bigskip

However, once an odd positive integer $N$ is known to be composite, it does
not appear to be an easy (polynomial time) task on a classical computer to
determine its prime factors. \ As mentioned earlier, so far the most
asymptotically efficient \emph{classical} algorithm known is the number
theoretic sieve \cite{Lenstra1}, \cite{Lenstra2}, which factors an integer $N
$ in time $O\left(  \exp\left[  \left(  \lg N\right)  ^{1/3}\left(  \lg\lg
N\right)  ^{2/3}\right]  \right)  $.

\bigskip

\noindent\textbf{Prime Factorization Problem.} \ \textit{Given a composite odd
positive integer }$N$\textit{, find its prime factors.}

\bigskip

It is well known\cite{Miller1} that factoring $N$ can be reduced to the task
of choosing at random an integer $m$ relatively prime to $N$, and then
determining its modulo $N$ multiplicative order $P$, i.e., to finding the
smallest positive integer $P$ such that
\[
m^{P}=1\operatorname{mod}N\text{ .}%
\]
It was precisely this approach to factoring that enabled Shor to construct his
factoring algorithm.

\vspace{0.5in}

\section{Overview of Shor's algorithm}

\quad\bigskip

But what is Shor's quantum factoring algorithm?

\vspace{0.5in}

Let $\mathbb{N}=\left\{  0,1,2,3,\ldots\right\}  $ denote the set of natural numbers.

\vspace{0.5in}

Shor's algorithm provides a solution to the above problem. \ His algorithm
consists of the five steps (\textbf{steps} \textbf{1} through \textbf{5}),
with only $\mathbb{STEP}$\textbf{\ 2} requiring the use of a quantum computer.
\ The remaining four other steps of the algorithm are to be performed on a
classical computer.

\bigskip

We begin by briefly describing all five steps. \ After that, we will then
focus in on the quantum part of the algorithm, i.e., $\mathbb{STEP}%
$\textbf{\ 2}.

\vspace{0.5in}

\begin{itemize}
\item [\fbox{\textbf{Step 1.}}]Choose a random positive integer $m$. \ Use the
polynomial time Euclidean algorithm\footnote{The Euclidean algorithm is
$O\left(  \lg^{2}N\right)  $. \ For a description of the Euclidean algorithm,
see for example \cite{Cox1} or \cite{Cormen1}.} to compute the greatest common
divisor $\gcd\left(  m,N\right)  $ of $m$ and $N$. \ If the greatest common
divisor $\gcd\left(  m,N\right)  \neq1$, then we have found a non-trivial
factor of $N$, and we are done. \ If, on the other hand, $\gcd\left(
m,N\right)  =1$, then proceed to $\mathbb{STEP}$\textbf{\ 2}.
\end{itemize}

\bigskip

\begin{itemize}
\item [\fbox{$\mathbb{STEP}$ \textbf{2.}}]Use a \textsc{quantum computer} to
determine the unknown period $P$ of the function
\[%
\begin{array}
[c]{ccc}%
\mathbb{N} & \overset{f_{N}}{\longrightarrow} & \mathbb{N}\\
a & \longmapsto &  m^{a}\operatorname{mod}N
\end{array}
\]
\end{itemize}

\bigskip

\begin{itemize}
\item [\fbox{\textbf{Step 3.}}]If $P$ is an odd integer, then goto
\textbf{Step 1}. \ [The probability of $P$ being odd is $(%
\frac12
)^{k}$, where $k$ is the number of distinct prime factors of $N$.] \ If $P$ is
even, then proceed to \textbf{Step 4}.
\end{itemize}

\bigskip\ 

\begin{itemize}
\item [\fbox{\textbf{Step 4.}}]Since $P$ is even,
\[
\left(  m^{P/2}-1\right)  \left(  m^{P/2}+1\right)  =m^{P}%
-1=0\operatorname{mod}N\text{ .}%
\]
If $m^{P/2}+1=0\operatorname{mod}N$, then goto \textbf{Step 1}. \ If
$m^{P/2}+1\neq0\operatorname{mod}N$, then proceed to \textbf{Step 5}. \ It can
be shown that the probability that $m^{P/2}+1=0\operatorname{mod}N$ is less
than $(%
\frac12
)^{k-1}$, where $k$ denotes the number of distinct prime factors of $N$.
\end{itemize}

\bigskip

\begin{itemize}
\item [\fbox{\textbf{Step 5.}}]Use the Euclidean algorithm to compute
$d=\gcd\left(  m^{P/2}-1,N\right)  $. \ Since $m^{P/2}+1\neq
0\operatorname{mod}N$, it can easily be shown that $d$ is a non-trivial factor
of $N$. \ Exit with the answer $d$.
\end{itemize}

\vspace{0.5in}

Thus, the task of factoring an odd positive integer $N$ reduces to the
following problem:

\bigskip

\noindent\textbf{Problem.} \textit{Given a periodic function }
\[
f:\mathbb{N}\longrightarrow\mathbb{N}\text{ ,}%
\]
\textit{find the period }$P$\textit{\ of }$f$\textit{.}

\vspace{0.5in}

\section{Preparations for the quantum part of Shor's algorithm}

\quad\bigskip

Choose a power of 2
\[
Q=2^{L}%
\]
such that
\[
N^{2}\leq Q=2^{L}<2N^{2}\text{ ,}%
\]
and consider $f$ restricted to the set
\[
S_{Q}=\left\{  0,1,\ldots,Q-1\right\}
\]
which we also denote by $f$, i.e.,
\[
f:S_{Q}\longrightarrow S_{Q}\text{ .}%
\]

\bigskip

In preparation for a discussion of $\mathbb{STEP}$ 2 of Shor's algorithm, we
construct two $L$-qubit quantum registers, \textsc{Register1} and
\textsc{Register2} to hold respectively the arguments and the values of the
function $f$, i.e.,
\[
\left|  \text{\textsc{Reg1}}\right\rangle \left|  \text{\textsc{Reg2}%
}\right\rangle =\left|  a\right\rangle \left|  f(a)\right\rangle =\left|
a\right\rangle \left|  b\right\rangle =\left|  a_{0}a_{1}\cdots a_{L-1}%
\right\rangle \left|  b_{0}b_{1}\cdots b_{L-1}\right\rangle
\]

In doing so, we have adopted the following convention for representing
integers in these registers:

\bigskip

\noindent\textbf{Notation Convention.} \ In a quantum computer, \textit{we
represent an integer }$a$\textit{\ with radix }$2$\textit{\ representation }
\[
a=\sum_{j=0}^{L-1}a_{j}2^{j}\text{ , }%
\]
\textit{as a quantum register consisting of the }$2^{n}$\textit{\ qubits }
\[
\left|  a\right\rangle =\left|  a_{0}a_{1}\cdots a_{L-1}\right\rangle =%
{\displaystyle\bigotimes\limits_{j=0}^{L-1}}
\left|  a_{j}\right\rangle
\]

\bigskip

For example, the integer $23$ is represented in our quantum computer as $n$
qubits in the state:
\[
\left|  23\right\rangle =\left|  10111000\cdots0\right\rangle
\]

\vspace{0.5in}

Before continuing, we remind the reader of the classical definition of the $Q
$-point Fourier transform.

\begin{definition}
Let $\omega$ be a primitive $Q$-th root of unity, e.g., $\omega=e^{2\pi i/Q}$.
Then the $Q$-point Fourier transform is the map
\begin{gather*}
Map(S_{Q},\mathbb{C})\overset{\mathcal{F}}{\longrightarrow}Map(S_{Q}%
,\mathbb{C})\\
\left[  f:S_{Q}\longrightarrow\mathbb{C}\right]  \longmapsto\left[
\widehat{f}:S_{Q}\longrightarrow\mathbb{C}\right]
\end{gather*}
where
\[
\widehat{f}\left(  y\right)  =\frac{1}{\sqrt{Q}}\sum_{x\in S_{Q}}%
f(x)\omega^{xy}%
\]
\end{definition}

\bigskip

We implement the Fourier transform $\mathcal{F}$\ as a unitary transformation,
which in the standard basis
\[
\left|  0\right\rangle ,\left|  1\right\rangle ,\ldots,\left|
Q-1\right\rangle
\]
is given by the $Q\times Q$ unitary matrix
\[
\mathcal{F}=\frac{1}{\sqrt{Q}}\left(  \omega^{xy}\right)  \text{ .}%
\]
This unitary transformation can be factored into the product of $O\left(
\lg^{2}Q\right)  =O\left(  \lg^{2}N\right)  $ sufficiently local unitary
transformations. (See \cite{Shor1}, \cite{Hoyer1}.)

\vspace{0.5in}

\section{The quantum part of Shor's algorithm}

\quad\bigskip

The quantum part of Shor's algorithm, i.e., $\mathbb{STEP}$ \textbf{2}, is the following:

\bigskip

\begin{itemize}
\item [\fbox{$\mathbb{STEP}$ \textbf{2.0}}]Initialize registers 1 and 2,
i.e.,
\[
\left|  \psi_{0}\right\rangle =\left|  \text{\textsc{Reg1}}\right\rangle
\left|  \text{\textsc{Reg2}}\right\rangle =\left|  0\right\rangle \left|
0\right\rangle =\left|  00\cdots0\right\rangle \left|  0\cdots0\right\rangle
\]

\item[\fbox{$\mathbb{STEP}$ \textbf{2.1}}] \footnote{In this step we could
have instead applied the Hadamard transform to \textsc{Register1} with the
same result, but at the computational cost of $O\left(  \lg N\right)  $
sufficiently local unitary transformations. \ The term sufficiently local
unitary transformationis defined in the last part of section 7.7 of
\cite{Lomonaco1}.}Apply the $Q$-point Fourier transform $\mathcal{F}$ to
\textsc{Register1}.
\[
\left|  \psi_{0}\right\rangle =\left|  0\right\rangle \left|  0\right\rangle
\overset{\mathcal{F}\otimes I}{\longmapsto}\left|  \psi_{1}\right\rangle
=\frac{1}{\sqrt{Q}}\sum_{x=0}^{Q-1}\omega^{0\cdot x}\left|  x\right\rangle
\left|  0\right\rangle =\frac{1}{\sqrt{Q}}\sum_{x=0}^{Q-1}\left|
x\right\rangle \left|  0\right\rangle
\]
\end{itemize}

\begin{remark}
Hence, \textsc{Register1} now holds all the integers
\[
0,1,2,\ldots,Q-1
\]
in superposition.
\end{remark}

\bigskip

\begin{itemize}
\item [\fbox{$\mathbb{STEP}$ \textbf{2.2}}]Let $U_{f}$ be the unitary
transformation that takes $\left|  x\right\rangle \left|  0\right\rangle $ to
$\left|  x\right\rangle \left|  f(x)\right\rangle $. \ Apply the linear
transformation $U_{f}$ to the two registers. \ The result is:
\[
\left|  \psi_{1}\right\rangle =\frac{1}{\sqrt{Q}}\sum_{x=0}^{Q-1}\left|
x\right\rangle \left|  0\right\rangle \overset{U_{f}}{\longmapsto}\left|
\psi_{2}\right\rangle =\frac{1}{\sqrt{Q}}\sum_{x=0}^{Q-1}\left|
x\right\rangle \left|  f(x)\right\rangle
\]
\end{itemize}

\bigskip

\begin{remark}
The state of the two registers is now more than a superposition of states.
\ In this step, we have quantum entangled the two registers.
\end{remark}

\bigskip

\bigskip

\begin{itemize}
\item [\fbox{$\mathbb{STEP}$ \textbf{2.3.}}]Apply the $Q$-point Fourier
transform $\mathcal{F}$ to \textsc{Reg1}. \ The resulting state is:
\[%
\begin{array}
[c]{ccl}%
\left|  \psi_{2}\right\rangle =\frac{1}{\sqrt{Q}}%
{\displaystyle\sum\limits_{x=0}^{Q-1}}
\left|  x\right\rangle \left|  f(x)\right\rangle  & \overset{\mathcal{F}%
\otimes I}{\longmapsto} & \left|  \psi_{3}\right\rangle =\frac{1}{Q}%
{\displaystyle\sum\limits_{x=0}^{Q-1}}
{\displaystyle\sum\limits_{y=0}^{Q-1}}
\omega^{xy}\left|  y\right\rangle \left|  f(x)\right\rangle \\
&  & \\
&  & \qquad=\frac{1}{Q}%
{\displaystyle\sum\limits_{y=0}^{Q-1}}
\left\|  \left|  \Upsilon(y)\right\rangle \right\|  \cdot\left|
y\right\rangle \frac{\left|  \Upsilon(y)\right\rangle }{\left\|  \left|
\Upsilon(y)\right\rangle \right\|  }\text{ ,}%
\end{array}
\]
where
\[
\left|  \Upsilon(y)\right\rangle =%
{\displaystyle\sum\limits_{x=0}^{Q-1}}
\omega^{xy}\left|  f(x)\right\rangle \text{. }%
\]
\end{itemize}

\bigskip

\begin{itemize}
\item [\fbox{$\mathbb{STEP}$ \textbf{2.4.}}]Measure \textsc{Reg1}, i.e.,
perform a measurement with respect to the orthogonal projections
\[
\left|  0\right\rangle \left\langle 0\right|  \otimes I,\ \left|
1\right\rangle \left\langle 1\right|  \otimes I,\ \left|  2\right\rangle
\left\langle 2\right|  \otimes I,\ \ldots\ ,\ \left|  Q-1\right\rangle
\left\langle Q-1\right|  \otimes I\text{ ,}%
\]
where $I$ denotes the identity operator on the Hilbert space of the second
register \textsc{Reg2}. \ 
\end{itemize}

\bigskip

As a result of this measurement, we have, with probability
\[
Prob\left(  y_{0}\right)  =\frac{\left\|  \left|  \Upsilon(y_{0})\right\rangle
\right\|  ^{2}}{Q^{2}}\text{ ,}%
\]
moved to the state
\[
\left|  y_{0}\right\rangle \frac{\left|  \Upsilon(y_{0})\right\rangle
}{\left\|  \left|  \Upsilon(y_{0})\right\rangle \right\|  }%
\]
and measured the value
\[
y_{0}\in\left\{  0,1,2,\ldots,Q-1\right\}  \text{ . }%
\]

\bigskip

If after this computation, we ignore the two registers \textsc{Reg1} and
\textsc{Reg2}, we see that what we have created is nothing more than a
classical probability distribution $\mathcal{S}$ on the sample space
\[
\left\{  0,1,2,\ldots,Q-1\right\}  \text{ .}%
\]
In other words, the sole purpose of executing STEPS 2.1 to 2.4 is to create a
classical finite memoryless stochastic source $\mathcal{S}$ which outputs a
symbol $y_{0}\in\left\{  0,1,2,\ldots,Q-1\right\}  $ with the probability
\[
Prob(y_{0})=\frac{\left\|  \left|  \Upsilon(y_{0})\right\rangle \right\|
^{2}}{Q^{2}}\text{ .}%
\]
(For more details, please refer to section 8.1 of \cite{Lomonaco1}.)

\bigskip

As we shall see, the objective of the remander of Shor's algorithm is to glean
information about the period $P$ of $f$ from the just created stochastic
source $\mathcal{S}$. The stochastic source was created exactly for that reason.

\bigskip

\section{Peter Shor's stochastic source $\mathcal{S}$}

\qquad\bigskip

Before continuing to the final part of Shor's algorithm, we need to analyze
the probability distribution $Prob\left(  y\right)  $ a little more carefully.

\bigskip

\begin{proposition}
Let $q$ and $r$ be the unique non-negative integers such that $Q=Pq+r$ , where
$0\leq r<P$ ; and let $Q_{0}=Pq$. \ Then
\[
Prob\left(  y\right)  =\left\{
\begin{array}
[c]{lrl}%
\frac{r\sin^{2}\left(  \frac{\pi Py}{Q}\cdot\left(  \frac{Q_{0}}{P}+1\right)
\right)  +\left(  P-r\right)  \sin^{2}\left(  \frac{\pi Py}{Q}\cdot\frac
{Q_{0}}{P}\right)  }{Q^{2}\sin^{2}\left(  \frac{\pi Py}{Q}\right)  } &
\text{if} & Py\neq0\operatorname{mod}Q\\
&  & \\
\frac{r\left(  Q_{0}+P\right)  ^{2}+\left(  P-r\right)  Q_{0}^{2}}{Q^{2}P^{2}}%
& \text{if} & Py=0\operatorname{mod}Q
\end{array}
\right.
\]
\end{proposition}

\begin{proof}
We begin by deriving a more usable expression for $\left|  \Upsilon
(y)\right\rangle $.
\[%
\begin{array}
[c]{rrl}%
\left|  \Upsilon(y)\right\rangle  & = &
{\displaystyle\sum\limits_{x=0}^{Q-1}}
\omega^{xy}\left|  f(x)\right\rangle =%
{\displaystyle\sum\limits_{x=0}^{Q_{0}-1}}
\omega^{xy}\left|  f(x)\right\rangle +%
{\displaystyle\sum\limits_{x=Q_{0}}^{Q-1}}
\omega^{xy}\left|  f(x)\right\rangle \\
&  & \\
& = &
{\displaystyle\sum\limits_{x_{0}=0}^{P-1}}
{\displaystyle\sum\limits_{x_{1}=0}^{\frac{Q_{0}}{P}-1}}
\omega^{\left(  Px_{1}+x_{0}\right)  y}\left|  f(Px_{1}+x_{0})\right\rangle +%
{\displaystyle\sum\limits_{x_{0}=0}^{r-1}}
\omega^{\left[  P\left(  \frac{Q_{0}}{P}\right)  +x_{0}\right]  y}\left|
f(Px_{1}+x_{0})\right\rangle \\
&  & \\
& = &
{\displaystyle\sum\limits_{x_{0}=0}^{P-1}}
\omega^{x_{0}y}\cdot\left(
{\displaystyle\sum\limits_{x_{1}=0}^{\frac{Q_{0}}{P}-1}}
\omega^{Pyx_{1}}\right)  \left|  f(x_{0})\right\rangle +%
{\displaystyle\sum\limits_{x_{0}=0}^{r-1}}
\omega^{x_{0}y}\cdot\omega^{Py\left(  \frac{Q_{0}}{P}\right)  }\left|
f(x_{0})\right\rangle \\
&  & \\
& = &
{\displaystyle\sum\limits_{x_{0}=0}^{r-1}}
\omega^{x_{0}y}\cdot\left(
{\displaystyle\sum\limits_{x_{1}=0}^{\frac{Q_{0}}{P}}}
\omega^{Pyx_{1}}\right)  \left|  f(x_{0})\right\rangle +%
{\displaystyle\sum\limits_{x_{0}=r}^{P-1}}
\omega^{x_{0}y}\cdot\left(
{\displaystyle\sum\limits_{x_{1}=0}^{\frac{Q_{0}}{P}-1}}
\omega^{Pyx_{1}}\right)  \left|  f(x_{0})\right\rangle
\end{array}
\]
where we have used the fact that $f$ is periodic of period $P$.

\bigskip

Since $f$ is one-to-one when restricted to its period $0,1,2,\ldots,P-1$, all
the kets
\[
\left|  f(0)\right\rangle ,\ \left|  f(1)\right\rangle ,\ \left|
f(2)\right\rangle ,\ \ldots\ ,\ \left|  f(P-1)\right\rangle ,\
\]
are mutually orthogonal. \ Hence,
\[
\left\langle \Upsilon(y)\mid\Upsilon(y)\right\rangle =r\left|
{\displaystyle\sum\limits_{x_{1}=0}^{\frac{Q_{0}}{P}}}
\omega^{Pyx_{1}}\right|  ^{2}+(P-r)\left|
{\displaystyle\sum\limits_{x_{1}=0}^{\frac{Q_{0}}{P}-1}}
\omega^{Pyx_{1}}\right|  ^{2}\text{ .}%
\]

\bigskip

If $Py=0\operatorname{mod}Q$, then since $\omega$ is a $Q$-th root of unity,
we have
\[
\left\langle \Upsilon(y)\mid\Upsilon(y)\right\rangle =r\left(  \frac{Q_{0}}%
{P}+1\right)  ^{2}+\left(  P-r\right)  \left(  \frac{Q_{0}}{P}\right)
^{2}\text{ .}%
\]

\bigskip

On the other hand, if $Py\neq0\operatorname{mod}Q$, then we can sum the
geometric series to obtain
\begin{align*}
\left\langle \Upsilon(y)\mid\Upsilon(y)\right\rangle  &  =r\left|
\frac{\omega^{Py\cdot\left(  \frac{Q_{0}}{P}+1\right)  }-1}{\omega^{Py}%
-1}\right|  ^{2}+\left(  P-r)\right)  \left|  \frac{\omega^{Py\cdot\left(
\frac{Q_{0}}{P}\right)  }-1}{\omega^{Py}-1}\right|  ^{2}\\
& \\
&  =r\left|  \frac{e^{\frac{2\pi i}{Q}\cdot Py\cdot\left(  \frac{Q_{0}}%
{P}+1\right)  }-1}{e^{\frac{2\pi i}{Q}\cdot Py}-1}\right|  ^{2}+\left(
P-r)\right)  \left|  \frac{e^{\frac{2\pi i}{Q}\cdot Py\cdot\left(  \frac
{Q_{0}}{P}\right)  }-1}{e^{\frac{2\pi i}{Q}\cdot Py}-1}\right|  ^{2}%
\end{align*}
where we have used the fact that $\omega$ is the primitive $Q$-th root of
unity given by
\[
\omega=e^{2\pi i/Q}\text{ .}%
\]

\bigskip

The remaining part of the proposition is a consequence of the trigonometric
identity
\[
\left|  e^{i\theta}-1\right|  ^{2}=4\sin^{2}\left(  \frac{\theta}{2}\right)
\text{ .}%
\]
\end{proof}

\bigskip

As a corollary, we have

\bigskip

\begin{corollary}
If $P$ is an exact divisor of $Q$, then
\[
Prob\left(  y\right)  =\left\{
\begin{array}
[c]{lrl}%
0 & \text{if} & Py\neq0\operatorname{mod}Q\\
&  & \\
\frac{1}{P} & \text{if} & Py=0\operatorname{mod}Q
\end{array}
\right.
\]
\end{corollary}

\bigskip

\section{A momentary digression: Continued fractions}

\qquad\bigskip

We digress for a moment to review the theory of continued fractions. (For a
more in-depth explanation of the theory of continued fractions, please refer
to \cite{Hardy1} and \cite{LeVeque1}.)

\bigskip

Every positive rational number $\xi$ can be written as an expression in the
form
\[
\xi=a_{0}+\frac{1}{a_{1}+\frac{\overset{}{\underset{}{1}}}{a_{2}%
+\frac{\overset{}{\underset{}{1}}}{a_{3}+\frac{\overset{}{\underset{}{1}}%
}{\cdots+\frac{\overset{}{\underset{}{1}}}{\overset{}{a_{N}}}}}}}\text{ ,}%
\]
where $a_{0}$ is a non-negative integer, and where $a_{1},\ldots,a_{N}$ are
positive integers. \ Such an expression is called a (finite, simple)
\textbf{continued fraction}, and is uniquely determined by $\xi$ provided we
impose the condition $a_{N}>1$. \ For typographical simplicity, we denote the
above continued fraction by
\[
\left[  a_{0},a_{1},\ldots,a_{N}\right]  \text{ .}%
\]

The continued fraction expansion of $\xi$ can be computed with the following
recurrence relation, which always terminates if $\xi$ is rational:
\[
\fbox{$\overset{}{\underset{}{%
\begin{array}
[c]{lll}%
\left\{
\begin{array}
[c]{r}%
a_{0}=\left\lfloor \xi\right\rfloor \\
\\
\xi_{0}=\xi-a_{0}%
\end{array}
\right.  \text{ ,} & \text{and if }\xi_{n}\neq0\text{, then} & \left\{
\begin{array}
[c]{l}%
a_{n+1}=\left\lfloor 1/\xi_{n}\right\rfloor \\
\\
\xi_{n+1}=\frac{1}{\xi_{n}}-a_{n+1}%
\end{array}
\right.
\end{array}
}}$}%
\]

\bigskip

The $n$-th \textbf{convergent} ($0\leq n\leq N$) of the above continued
fraction is defined as the rational number $\xi_{n}$ given by
\[
\xi_{n}=\left[  a_{0},a_{1},\ldots,a_{n}\right]  \text{ .}%
\]
Each convergent $\xi_{n}$ can be written in the form, $\xi_{n}=\frac{p_{n}%
}{q_{n}}$, where $p_{n}$ and $q_{n}$ are relatively prime integers (
$\gcd\left(  p_{n},q_{n}\right)  =1$). The integers $p_{n}$ and $q_{n}$ are
determined by the recurrence relation
\[
\fbox{$%
\begin{array}
[c]{lll}%
p_{0}=a_{0}, & p_{1}=a_{1}a_{0}+1, & p_{n}=a_{n}p_{n-1}+p_{n-2},\\
&  & \\
q_{0}=1, & q_{1}=a_{1}, & q_{n}=a_{n}q_{n-1}+q_{n-2}\text{ \ .}%
\end{array}
$}%
\]

\bigskip

\bigskip

\section{Preparation for the final part of Shor's algorithm}

\qquad\bigskip

\begin{definition}
\footnote{$\left\{  a\right\}  _{Q}=a-Q\cdot round\left(  \frac{a}{Q}\right)
=a-Q\cdot\left\lfloor \frac{a}{Q}+\frac{1}{2}\right\rfloor $.}For each integer
$\ \ a$, let $\left\{  a\right\}  _{Q}$ denote the \textbf{residue} of
$\ \ a\ \ $ modulo $Q$ \textbf{of smallest magnitude}. \ In other words,
$\left\{  a\right\}  _{Q}$ is the unique integer such that
\[
\left\{
\begin{array}
[c]{l}%
a=\left\{  a\right\}  _{Q}\operatorname{mod}Q\\
\\
-Q/2<\left\{  a\right\}  _{Q}\leq Q/2
\end{array}
\right.  \text{ .}%
\]
\end{definition}

\bigskip

\begin{proposition}
Let $y$ be an integer lying in $S_{Q}$. \ Then
\[
Prob\left(  y\right)  \geq\left\{
\begin{array}
[c]{lrl}%
\frac{4}{\pi^{2}}\cdot\frac{1}{P}\cdot\left(  1-\frac{1}{N}\right)  ^{2} &
\text{if} & 0<\left|  \left\{  Py\right\}  _{Q}\right|  \leq\frac{P}{2}%
\cdot\left(  1-\frac{1}{N}\right) \\
&  & \\
\frac{1}{P}\cdot\left(  1-\frac{1}{N}\right)  ^{2} & \text{if} & \left\{
Py\right\}  _{Q}=0
\end{array}
\right.
\]
\end{proposition}

\begin{proof}
We begin by noting that
\[%
\begin{array}
[c]{ll}%
\left|  \frac{\pi\left\{  Py\right\}  _{Q}}{Q}\cdot\left(  \frac{Q_{0}}%
{P}+1\right)  \right|  & \leq\frac{\pi}{Q}\cdot\frac{P}{2}\cdot\left(
1-\frac{1}{N}\right)  \cdot\left(  \frac{Q_{0}+P}{P}\right)  \leq\frac{\pi}%
{2}\cdot\left(  1-\frac{1}{N}\right)  \cdot\left(  \frac{Q+P}{Q}\right) \\
& \\
& \leq\frac{\pi}{2}\cdot\left(  1-\frac{1}{N}\right)  \cdot\left(  1+\frac
{P}{Q}\right)  \leq\frac{\pi}{2}\cdot\left(  1-\frac{1}{N}\right)
\cdot\left(  1+\frac{N}{N^{2}}\right)  <\frac{\pi}{2}\text{ ,}%
\end{array}
\]
where we have made use of the inequalities
\[
N^{2}\leq Q<2N^{2}\text{ \ and \ }0<P\leq N\text{ \ .}%
\]
It immediately follows that
\[
\left|  \frac{\pi\left\{  Py\right\}  _{Q}}{Q}\cdot\frac{Q_{0}}{P}\right|
<\frac{\pi}{2}\text{ \ .}%
\]

\bigskip

As a result, we can legitimately use the inequality
\[
\frac{4}{^{\pi^{2}}}\theta^{2}\leq\sin^{2}\theta\leq\theta^{2}\text{, for
}\left|  \theta\right|  <\frac{\pi}{2}%
\]
to simplify the expression for $Prob\left(  y\right)  $.

\bigskip

Thus,
\[%
\begin{array}
[c]{lll}%
Prob\left(  y\right)  & = & \frac{r\sin^{2}\left(  \frac{\pi\left\{
Py\right\}  _{Q}}{Q}\cdot\left(  \frac{Q_{0}}{P}+1\right)  \right)  +\left(
P-r\right)  \sin^{2}\left(  \frac{\pi\left\{  Py\right\}  _{Q}}{Q}\cdot
\frac{Q_{0}}{P}\right)  }{Q^{2}\sin^{2}\left(  \frac{\pi Py}{Q}\right)  }\\
&  & \\
& \geq & \frac{r\cdot\frac{4}{\pi^{2}}\cdot\left(  \frac{\pi\left\{
Py\right\}  _{Q}}{Q}\cdot\left(  \frac{Q_{0}}{P}+1\right)  \right)
^{2}+\left(  P-r\right)  \cdot\frac{4}{\pi^{2}}\cdot\left(  \frac{\pi\left\{
Py\right\}  _{Q}}{Q}\cdot\frac{Q_{0}}{P}\right)  ^{2}}{Q^{2}\left(  \frac
{\pi\left\{  Py\right\}  _{Q}}{Q}\right)  ^{2}}\\
&  & \\
& \geq & \frac{4}{\pi^{2}}\cdot\frac{P\cdot\left(  \frac{Q_{0}}{P}\right)
^{2}}{Q^{2}}=\frac{4}{\pi^{2}}\cdot\frac{1}{P}\cdot\left(  \frac{Q-r}%
{Q}\right)  ^{2}\\
&  & \\
& = & \frac{4}{\pi^{2}}\cdot\frac{1}{P}\cdot\left(  1-\frac{r}{Q}\right)
^{2}\geq\frac{4}{\pi^{2}}\cdot\frac{1}{P}\cdot\left(  1-\frac{1}{N}\right)
^{2}%
\end{array}
\]

\bigskip

The remaining case, $\left\{  Py\right\}  _{Q}=0$ is left to the reader.
\end{proof}

\bigskip

\begin{lemma}
Let
\[
Y=\left\{  y\in S_{Q}\mid\left|  \left\{  Py\right\}  _{Q}\right|  \leq
\frac{P}{2}\right\}  \text{ \ \ and \ \ }S_{P}=\left\{  d\in S_{Q}\mid0\leq
d<P\right\}  \text{ .}%
\]
Then the map
\[%
\begin{array}
[c]{lll}%
Y & \longrightarrow &  S_{P}\\
y & \longmapsto &  d=d(y)=round\left(  \frac{P}{Q}\cdot y\right)
\end{array}
\]
is a bijection with inverse
\[
y=y(d)=round\left(  \frac{Q}{P}\cdot d\right)  \text{ .}%
\]
Hence, $Y$ and $S_{P}$ are in one-to-one correspondence. \ Moreover,
\[
\left\{  Py\right\}  _{Q}=P\cdot y-Q\cdot d(y)\text{ .}%
\]
\end{lemma}

\bigskip

\begin{remark}
Moreover, the following two sets of rationals are in one-to-one
correspondence
\[
\left\{  \frac{y}{Q}\mid y\in Y\right\}  \longleftrightarrow\left\{  \frac
{d}{P}\mid0\leq d<P\right\}
\]
\end{remark}

\bigskip

As a result of the measurement performed in $\mathbb{STEP}$ 2.4, we have in
our possession an integer $y\in Y$. \ We now show how $y$ \ can be use to
determine the unknown period $P$. \ 

\bigskip

We now need the following theorem\footnote{See \cite[Theorem 184, Section
10.15]{Hardy1}.} from the theory of continued fractions:

\bigskip

\begin{theorem}
Let $\xi$ be a real number, and let $a$ and $b$ be integers with $b>0$. If
\[
\left|  \xi-\frac{a}{b}\right|  \leq\frac{1}{2b^{2}}\text{ ,}%
\]
then the rational number $a/b$ is a convergent of the continued fraction
expansion of $\xi$. \ 
\end{theorem}

\bigskip

As a corollary, we have:

\bigskip

\begin{corollary}
If $\left|  \left\{  Py\right\}  _{Q}\right|  \leq\frac{P}{2}$, then the
rational number $\frac{d(y)}{P}$ is a convergent of the continued fraction
expansion of $\frac{y}{Q}$. \ 
\end{corollary}

\begin{proof}
Since
\[
Py-Qd(y)=\left\{  Py\right\}  _{Q}\text{ ,}%
\]
we know that
\[
\left|  Py-Qd(y)\right|  \leq\frac{P}{2}\text{, }%
\]
which can be rewritten as
\[
\left|  \frac{y}{Q}-\frac{d(y)}{P}\right|  \leq\frac{1}{2Q}\text{ .}%
\]
But, since $Q\geq N^{2}$, it follows that
\[
\left|  \frac{y}{Q}-\frac{d(y)}{P}\right|  \leq\frac{1}{2N^{2}}\text{ .}%
\]
Finally, since $P\leq N$ (and hence $\frac{1}{2N^{2}}\leq\frac{1}{2P^{2}})$,
the above theorem can be applied. \ Thus, $\frac{d(y)}{P}$ is a convergent of
the continued fraction expansion of $\xi=\frac{y}{Q}$. \ 
\end{proof}

\bigskip

Since $\frac{d(y)}{P}$ is a convergent of the continued fraction expansion of
$\frac{y}{Q}$, it follows that, for some $n$,
\[
\frac{d(y)}{P}=\frac{p_{n}}{q_{n}}\text{ ,}%
\]
where $p_{n}$ and $q_{n}$ are relatively prime positive integers given by a
recurrence relation found in the previous subsection. \ So it would seem that
we have found a way of deducing the period $P$ from the output $y$ of
$\mathbb{STEP}$ 2.4, and so we are done. \ 

\bigskip

Not quite! \ 

\bigskip

We can determine $P$ from the measured $y$ produced by $\mathbb{STEP}$ 2.4,
only if
\[
\left\{
\begin{array}
[c]{l}%
p_{n}=d(y)\\
\\
q_{n}=P
\end{array}
\right.  \text{ ,}%
\]
which is true only when $d(y)$ and $P$ are relatively prime.

\bigskip

So what is the probability that the $y\in Y$ produced by $\mathbb{STEP}$ 2.4
satisfies the additional condition that
\[
\gcd\left(  P,d(y)\right)  =1\text{ ?}%
\]

\bigskip

\begin{proposition}
The probability that the random $y$ produced by $\mathbb{STEP}$ 2.4 is such
that $d(y)$ and $P$ are relatively prime is bounded below by the following
expression
\[
Prob\left\{  y\in Y\mid\gcd(d(y),P)=1\right\}  \geq\frac{4}{\pi^{2}}\cdot
\frac{\phi(P)}{P}\cdot\left(  1-\frac{1}{N}\right)  ^{2}\text{ ,}%
\]
where $\phi(P)$ denotes Euler's totient function, i.e., $\phi(P)$ is the
number of positive integers less than $P$ which are relatively prime to $P$. \ 
\end{proposition}

\bigskip

The following theorem can be found in \cite[Theorem 328, Section 18.4]{Hardy1}:

\bigskip

\begin{theorem}%
\[
\lim\inf\frac{\phi(N)}{N/\ln\ln N}=e^{-\gamma}\text{,}%
\]
where $\gamma$ denotes Euler's constant $\gamma=0.57721566490153286061\ldots$
, and where $e^{-\gamma}=0.5614594836\ldots$ . \ 
\end{theorem}

\bigskip

As a corollary, we have:

\bigskip

\begin{corollary}%
\[
Prob\left\{  y\in Y\mid\gcd(d(y),P)=1\right\}  \geq\frac{4}{\pi^{2}\ln2}%
\cdot\frac{e^{-\gamma}-\epsilon\left(  P\right)  }{\lg\lg N}\cdot\left(
1-\frac{1}{N}\right)  ^{2}\text{ ,}%
\]
where $\epsilon\left(  P\right)  $ is a monotone decreasing sequence
converging to zero. \ In terms of asymptotic notation,
\[
Prob\left\{  y\in Y\mid\gcd(d(y),P)=1\right\}  =\Omega\left(  \frac{1}{\lg\lg
N}\right)  \text{ .}%
\]
Thus, if $\ \mathbb{STEP}$ 2.4 is repeated $O(\lg\lg N)$ times, then the
probability of success is $\Omega\left(  1\right)  $.
\end{corollary}

\begin{proof}
From the above theorem, we know that
\[
\frac{\phi(P)}{P/\ln\ln P}\geq e^{-\gamma}-\epsilon\left(  P\right)  \text{ .}%
\]
where $\epsilon\left(  P\right)  $ is a monotone decreasing sequence of
\ positive reals converging to zero. \ Thus,
\[
\frac{\phi(P)}{P}\geq\frac{e^{-\gamma}-\epsilon\left(  P\right)  }{\ln\ln
P}\geq\frac{e^{-\gamma}-\epsilon\left(  P\right)  }{\ln\ln N}=\frac
{e^{-\gamma}-\epsilon\left(  P\right)  }{\ln\ln2+\ln\lg N}\geq\frac
{e^{-\gamma}-\epsilon\left(  P\right)  }{\ln2}\cdot\frac{1}{\lg\lg N}%
\]
\end{proof}

\bigskip

\begin{remark}
$\Omega(\frac{1}{\lg\lg N})$ denotes an asymptotic lower bound. \ Readers not
familiar with the big-oh $O(\ast)$ and big-omega $\Omega\left(  \ast\right)  $
notation should refer to \cite[Chapter 2]{Cormen1} or \cite[Chapter
2]{Brassard1}.
\end{remark}

\bigskip

\begin{remark}
For the curious reader, lower bounds $LB(P)$ of $e^{-\gamma}-\epsilon\left(
P\right)  $ for $3\leq P\leq841$ are given in the following table:
\[%
\begin{tabular}
[c]{|l||l|}\hline\hline
\multicolumn{1}{||l||}{$P$} & \multicolumn{1}{||l||}{$LB(P)$}\\\hline\hline
3 & 0.062\\\hline
4 & 0.163\\\hline
5 & 0.194\\\hline
7 & 0.303\\\hline
13 & 0.326\\\hline
31 & 0.375\\\hline
61 & 0.383\\\hline
211 & 0.411\\\hline
421 & 0.425\\\hline
631 & 0.435\\\hline
841 & 0.468\\\hline
\end{tabular}
\]
Thus, if one wants a reasonable bound on the $Prob\left\{  y\in Y\mid
\gcd(d(y),P)=1\right\}  $ before continuing with Shor's algorithm, it would
pay to first use a classical algorithm to verify that the period $P$ of the
randomly chosen integer $m$ is not too small.
\end{remark}

\bigskip

\section{The final part of Shor's algorithm}

\qquad\bigskip

We are now prepared to give the last step in Shor's algorithm. \ This step can
be performed on a classical computer.

\bigskip

\begin{itemize}
\item [\fbox{\textbf{Step 2.5}}]Compute the period $P$ from the integer $y$
produced by $\mathbb{STEP}$ 2.4.
\end{itemize}

\bigskip

\begin{itemize}
\item
\begin{itemize}
\item [\qquad]\textsc{Loop} \textsc{for each} $n$ \textsc{from} $n=1$
\textsc{Until} $\xi_{n}=0$.
\end{itemize}
\end{itemize}

\bigskip

\begin{itemize}
\item
\begin{itemize}
\item
\begin{itemize}
\item [\qquad]Use the recurrence relations given in subsection 13.7, to
compute the $p_{n}$ and $q_{n}$ of the $n$-th convergent $\frac{p_{n}}{q_{n}}
$ of $\frac{y}{Q}$.
\end{itemize}
\end{itemize}
\end{itemize}

\bigskip

\begin{itemize}
\item
\begin{itemize}
\item
\begin{itemize}
\item [\qquad]Test to see if $q_{n}=P$ by computing\footnote{The indicated
algorithm for computing $m^{q_{n}}\operatorname{mod}N$ requires $O(\lg q_{n})$
arithmetic operations.}
\[
m^{q_{n}}=%
{\displaystyle\prod\limits_{i}}
\left(  m^{2^{i}}\right)  ^{q_{n,i}}\operatorname{mod}N\text{ ,}%
\]
where $q_{n}=\sum_{i}q_{n,i}2^{i}$ is the binary expansion of $q_{n}$.

\item[\qquad] If $m^{q_{n}}=1\operatorname{mod}N$, then exit with the answer
$P=q_{n}$, and proceed to \textbf{Step 3}. \ If not, then continue the loop.
\end{itemize}
\end{itemize}
\end{itemize}

\bigskip

\begin{itemize}
\item
\begin{itemize}
\item [\qquad]\textsc{End of Loop}
\end{itemize}
\end{itemize}

\bigskip

\begin{itemize}
\item
\begin{itemize}
\item [\qquad]If you happen to reach this point, you are a very unlucky
quantum computer scientist. \ You must start over by returning to
$\mathbb{STEP}$ 2.0. \ But don't give up hope! \ The probability that the
integer $y$ produced by $\mathbb{STEP}$ 2.4 will lead to a successful
completion of Step 2.5 is bounded below by
\[
\frac{4}{\pi^{2}\ln2}\cdot\frac{e^{-\gamma}-\epsilon\left(  P\right)  }{\lg\lg
N}\cdot\left(  1-\frac{1}{N}\right)  ^{2}>\frac{0.232}{\lg\lg N}\cdot\left(
1-\frac{1}{N}\right)  ^{2}\text{ ,}%
\]
provided the period $P$ is greater than $3$. \ [ $\gamma$ denotes Euler's constant.]
\end{itemize}
\end{itemize}

\bigskip

\section{\textbf{\bigskip}An example of Shor's algorithm}

\quad\bigskip

Let us now show how $N=91\ (=7\cdot13)$ can be factored using Shor's algorithm.

\bigskip

We choose $Q=2^{14}=16384$ so that $N^{2}\leq Q<2N^{2}$.

\bigskip

\begin{itemize}
\item [\fbox{\textbf{Step 1}}]Choose a random positive integer $m$, say $m=3$.
\ Since $\gcd(91,3)=1$, we proceed to $\mathbb{STEP}$\textbf{\ 2} to find the
period of the function $f$ given by \
\[
f(a)=3^{a}\operatorname{mod}91
\]
\end{itemize}

\begin{remark}
Unknown to us, $f$ has period $P=6$. For,
\[%
\begin{tabular}
[c]{||l||}\hline\hline
$%
\begin{array}
[c]{ccccccccccc}%
a &  & 0 & 1 & 2 & 3 & 4 & 5 & 6 & 7 & \cdots\\
&  &  &  &  &  &  &  &  &  & \\
f(a) &  & 1 & 3 & 9 & 27 & 81 & 61 & 1 & 3 & \cdots
\end{array}
$\\\hline\hline
\multicolumn{1}{||c||}{$\therefore\text{ Unknown period }P=6$}\\\hline\hline
\end{tabular}
\]
\end{remark}

\vspace{0.5in}

\begin{itemize}
\item [\fbox{$\mathbb{STEP}$ \textbf{2.0}}]Initialize registers 1 and 2. Thus,
the state of the two registers becomes:
\[
\left|  \psi_{0}\right\rangle =\left|  0\right\rangle \left|  0\right\rangle
\]
\end{itemize}

\vspace{0.5in}

\begin{itemize}
\item [\fbox{$\mathbb{STEP}$ \textbf{2.1}}]Apply the $Q$-point Fourier
transform $\mathcal{F}$ to register \#1, where
\[
\mathcal{F}\left|  k\right\rangle =\frac{1}{\sqrt{16384}}\sum_{x=0}%
^{16383}\omega^{0\cdot x}\left|  x\right\rangle \text{ ,}%
\]
and where $\omega$ is a primitive $Q$-th root of unity, e.g., $\omega
=e^{\frac{2\pi i}{16384}}$. Thus the state of the two registers becomes:
\[
\left|  \psi_{1}\right\rangle =\frac{1}{\sqrt{16384}}\sum_{x=0}^{16383}\left|
x\right\rangle \left|  0\right\rangle
\]
\end{itemize}

\vspace{0.5in}

\begin{itemize}
\item [\fbox{$\mathbb{STEP}$ \textbf{2.2}}]Apply the unitary transformation
$U_{f}$ to registers \#1 and \#2, where
\[
U_{f}\left|  x\right\rangle \left|  \ell\right\rangle =\left|  x\right\rangle
\left|  \ f(x)-\ell\ \operatorname{mod}91\right\rangle \text{ .}%
\]
(Please note that $U_{f}^{2}=I$.) Thus, the state of the two registers
becomes:
\[%
\begin{array}
[c]{rrrl}%
\left|  \psi_{2}\right\rangle  & = & \frac{1}{\sqrt{16384}} & \sum
_{x=0}^{16383}\left|  x\right\rangle \left|  3^{x}\operatorname{mod}%
91\right\rangle \\
&  &  & \\
& = & \frac{1}{\sqrt{16384}}( & \quad\left|  \ 0\right\rangle \left|
1\right\rangle \ +\left|  \ 1\right\rangle \left|  3\right\rangle +\left|
\ 2\right\rangle \left|  9\right\rangle \ +\left|  \ 3\right\rangle \left|
27\right\rangle +\left|  \ 4\right\rangle \left|  81\right\rangle +\left|
\ 5\right\rangle \left|  61\right\rangle \\
&  &  & \\
&  &  & +\ \left|  \ 6\right\rangle \left|  1\right\rangle \ +\left|
\ 7\right\rangle \left|  3\right\rangle +\left|  \ 8\right\rangle \left|
9\right\rangle \ +\left|  \ 9\right\rangle \left|  27\right\rangle +\left|
10\right\rangle \left|  81\right\rangle +\left|  11\right\rangle \left|
61\right\rangle \\
&  &  & \\
&  &  & +\ \left|  12\right\rangle \left|  1\right\rangle \ +\left|
13\right\rangle \left|  3\right\rangle \ +\left|  14\right\rangle \left|
9\right\rangle \ +\left|  15\right\rangle \left|  27\right\rangle +\left|
16\right\rangle \left|  81\right\rangle +\left|  17\right\rangle \left|
61\right\rangle \\
&  &  & \\
&  &  & +\ \ldots\\
&  &  & \\
&  &  & +\ \left|  16380\right\rangle \left|  1\right\rangle +\left|
16381\right\rangle \left|  3\right\rangle +\left|  16382\right\rangle \left|
9\right\rangle +\left|  16383\right\rangle \left|  27\right\rangle \\
&  & ) &
\end{array}
\]
\end{itemize}

\bigskip

\begin{remark}
The state of the two registers is now more than a superposition of states.
\ We have in the above step quantum entangled the two registers.
\end{remark}

\vspace{0.5in}

\begin{itemize}
\item [\fbox{$\mathbb{STEP}$ \textbf{2.3}}]Apply the $Q$-point $\mathcal{F} $
again to register \#1. Thus, the state of the system becomes:
\[%
\begin{array}
[c]{rrl}%
\left|  \psi_{3}\right\rangle  & = & \frac{1}{\sqrt{16384}}\sum_{x=0}%
^{16383}\frac{1}{\sqrt{16384}}\sum_{y=0}^{16383}\omega^{xy}\left|
y\right\rangle \left|  3^{x}\operatorname{mod}91\right\rangle \\
&  & \\
& = & \frac{1}{16384}\sum_{x=0}^{16383}\left|  y\right\rangle \sum
_{x=0}^{16383}\omega^{xy}\left|  3^{x}\operatorname{mod}91\right\rangle \\
&  & \\
& = & \frac{1}{16384}\sum_{x=0}^{16383}\left|  y\right\rangle \left|
\Upsilon\left(  y\right)  \right\rangle \text{ ,}%
\end{array}
\]
where
\[
\left|  \Upsilon\left(  y\right)  \right\rangle =\sum_{x=0}^{16383}\omega
^{xy}\left|  3^{x}\operatorname{mod}91\right\rangle
\]
Thus,
\[%
\begin{array}
[c]{rl}%
\left|  \Upsilon\left(  y\right)  \right\rangle = & \quad\quad\ \ \ \left|
1\right\rangle \ +\ \ \omega^{y}\left|  3\right\rangle +\ \omega^{2y}\left|
9\right\rangle \ +\ \omega^{3y}\left|  27\right\rangle +\ \ \omega^{4y}\left|
81\right\rangle +\ \ \omega^{5y}\left|  61\right\rangle \\
& \\
& +\ \ \omega^{6y}\left|  1\right\rangle \ +\ \omega^{7y}\left|
3\right\rangle +\ \omega^{8y}\left|  9\right\rangle \ +\ \omega^{9y}\left|
27\right\rangle +\ \omega^{10y}\left|  81\right\rangle +\omega^{11y}\left|
61\right\rangle \\
& \\
& +\ \omega^{12y}\left|  1\right\rangle \ +\omega^{13y}\left|  3\right\rangle
+\omega^{14y}\left|  9\right\rangle \ +\omega^{15y}\left|  27\right\rangle
+\omega^{16y}\left|  81\right\rangle +\omega^{17y}\left|  61\right\rangle \\
& \\
& +\ \ldots\\
& \\
& +\ \omega^{16380y}\left|  1\right\rangle +\omega^{16381y}\left|
3\right\rangle +\omega^{16382y}\left|  9\right\rangle +\omega^{16383y}\left|
27\right\rangle
\end{array}
\]
\end{itemize}

\vspace{0.5in}

\begin{itemize}
\item [\fbox{$\mathbb{STEP}$ \textbf{2.4}}]Measure \textsc{Reg1}. \ The result
of our measurement just happens to turn out to be
\[
y=13453
\]
\end{itemize}

\bigskip

Unknown to us, the probability of obtaining this particular $y$ is: \ \
\[
0.3189335551\times10^{-6}\text{ . }%
\]
Moreover, unknown to us, we're lucky! \ The corresponding $d$ is relatively
prime to $P$, i.e.,
\[
d=d(y)=round(\frac{P}{Q}\cdot y)=5
\]

\bigskip

However, we do know that the probability of $d(y)$ being relatively prime to
$P$ is greater than
\[
\frac{0.232}{\lg\lg N}\cdot\left(  1-\frac{1}{N}\right)  ^{2}\thickapprox
8.4\%\text{ \ (provided }P>3\text{),}%
\]
and we also know that
\[
\frac{d(y)}{P}%
\]
is a convergent of the continued fraction expansion of
\[
\xi=\frac{y}{Q}=\frac{13453}{16384}%
\]

So with a reasonable amount of confidence, we proceed to \textbf{Step 2.5}.

\bigskip

\begin{itemize}
\item [\fbox{\textbf{Step 2.5}}]Using the recurrence relations found in
subsection 13.7 of this paper, we successively compute (beginning with $n=0$)
the $a_{n}$'s and $q_{n}$'s for the continued fraction expansion of
\[
\xi=\frac{y}{Q}=\frac{13453}{16384}\text{ .}%
\]
For each non-trivial $n$ in succession, we check to see if
\[
3^{q_{n}}=1\operatorname{mod}91\text{. }%
\]
If this is the case, then we know $q_{n}=P$, and we immediately exit from
\textbf{Step 2.5} and proceed to \textbf{Step 3}.
\end{itemize}

\bigskip

\begin{itemize}
\item  In this example, $n=0$ and $n=1$ are trivial cases. \ 
\end{itemize}

\bigskip

\begin{itemize}
\item  For $n=2$, $a_{2}=4$ and $q_{2}=5$ . \ We test $q_{2}$ by computing
\[
3^{q_{2}}=3^{5}=\left(  3^{2^{0}}\right)  ^{1}\cdot\left(  3^{2^{1}}\right)
^{0}\cdot\left(  3^{2^{0}}\right)  ^{1}=61\neq1\operatorname{mod}91\text{ .}%
\]
Hence, $q_{2}\neq P$.
\end{itemize}

\bigskip

\begin{itemize}
\item  We proceed to $n=3$, and compute
\[
a_{3}=1\text{ and }q_{3}=6\text{. }%
\]
We then test $q_{3}$ by computing
\[
3^{q_{3}}=3^{6}=\left(  3^{2^{0}}\right)  ^{0}\cdot\left(  3^{2^{1}}\right)
^{1}\cdot\left(  3^{2^{0}}\right)  ^{1}=1\operatorname{mod}91\text{ .}%
\]
Hence, $q_{3}=P$. \ Since we now know the period $P$, there is no need to
continue to compute the remaining $a_{n}$'s and $q_{n}$'s. \ We proceed
immediately to \textbf{Step 3}.
\end{itemize}

\bigskip

To satisfy the reader's curiosity we have listed in the table below all the
values of $a_{n}$, $p_{n}$, and $q_{n}$ for $n=0,1,\ldots,14$. \ But it should
be mentioned again that we need only to compute $a_{n}$ and $q_{n}$ for
$n=0,1,2,3$, as indicated above. \
\[%
\begin{tabular}
[c]{|c||r|r|r|r|r|r|r|r|r|r|r|r|r|r|r|}\hline
$n$ & 0 & 1 & 2 & \textbf{3} & 4 & 5 & 6 & 7 & 8 & 9 & 10 & 11 & 12 & 13 &
14\\\hline\hline
$a_{n}$ & 0 & 1 & 4 & \textbf{1} & 1 & 2 & 3 & 1 & 1 & 3 & 1 & 1 & 1 & 1 &
3\\\hline
$p_{n}$ & 0 & 1 & 4 & \textbf{5} & 9 & 23 & 78 & 101 & 179 & 638 & 817 &
1455 & 2272 & 3727 & 13453\\\hline
$q_{n}$ & 1 & 1 & 5 & \textbf{6} & 11 & 28 & 95 & 123 & 218 & 777 & 995 &
1772 & 2767 & 4539 & 16384\\\hline
\end{tabular}
\]

\bigskip

\begin{itemize}
\item [\fbox{\textbf{Step 3.}}]Since $P=6$ is even, we proceed to \textbf{Step
4}.
\end{itemize}

\bigskip

\begin{itemize}
\item [\fbox{\textbf{Step 4.}}]Since
\[
3^{P/2}=3^{3}=27\neq-1\operatorname{mod}91\text{, }%
\]
we goto \textbf{Step 5}. \ 
\end{itemize}

\bigskip

\begin{itemize}
\item [\fbox{\textbf{Step 5.}}]With the Euclidean algorithm, we compute
\[
\gcd\left(  3^{P/2}-1,91\right)  =\gcd\left(  3^{3}-1,91\right)  =\gcd\left(
26,91\right)  =13\text{ .}%
\]
We have succeeded in finding a non-trivial factor of $N=91$, namely $13$. \ We
exit Shor's algorithm, and proceed to celebrate!
\end{itemize}

\bigskip

\bigskip

\begin{thebibliography}{99}
\bibitem{Brassard1}Brassard, Gilles, and Paul Bratley, ``\textbf{Algorithmics:
Theory and Practice,''} Printice-Hall, (1988).

\bibitem{Cormen1}Cormen, Thomas H., Charles E. Leiserson, and Ronald L.
Rivest, ``\textbf{Introduction to Algorithms},'' McGraw-Hill, (1990).

\bibitem{Cox1}Cox, David, John Little, and Donal O'Shea, ``\textbf{Ideals,
Varieties, and Algorithms},'' (second edition), Springer-Verlag, (1996).

\bibitem{Ekert1}Ekert, Artur K.and Richard Jozsa, \textbf{Quantum computation
and Shor's factoring algorithm}, Rev. Mod. Phys., 68,(1996), pp 733-753.

\bibitem{Hardy1}Hardy, G.H., and E.M. Wright, ``\textbf{An Introduction to the
Theory of Numbers},'' Oxford Press, (1965).

\bibitem{Hoyer1}Hoyer, Peter, \textbf{Efficient quantum transforms}, quant-ph/9702028.

\bibitem{Jozsa2}Jozsa, Richard, \textbf{Quantum algorithms and the Fourier
transform}, quant-ph preprint archive 9707033 17 Jul 1997.

\bibitem{Jozsa3}Jozsa, Richard, Proc. Roy. Soc. London Soc., Ser. A, 454,
(1998), 323 - 337.

\bibitem{Kitaev1}Kitaev, A., \textbf{Quantum measurement and the abelian
stabiliser problem,} (1995), quant-ph preprint archive 9511026.

\bibitem{Lenstra1}Lenstra, A.K., and H.W. Lenstra, Jr., eds., ``\textbf{The
Development of the Number Field Sieve},'' Lecture Notes in Mathematics, Vol.
1554, Springer-Velag, (1993).

\bibitem{Lenstra2}Lenstra, A.K., H.W. Lenstra, Jr., M.S. Manasse, and J.M.
Pollard, \textbf{The number field sieve}. Proc. 22nd Annual ACM\ Symposium on
Theory of ComputingACM, New York, (1990), \ pp 564 - 572. \ (See exanded
version in Lenstra \& Lenstra, (1993), pp 11 - 42.)

\bibitem{LeVeque1}LeVeque, William Judson, ``\textbf{Topics in Number Theory:
Volume I},'' Addison-Wesley, (1958).

\bibitem{Lomonaco1}Lomonaco, Samuel J., Jr., \textbf{A Rosetta Stone for
quantum mechanics with an introduction to quantum computation: Lecture Notes
for the AMS Short Course on Quantum Computation, Washington, DC, January
2000,} in \textbf{``Quantum Computation,''} edited by S.J. Lomonaco, Jr.,
AMS\ PSAPM Series. (to appear)

\bibitem{Miller1}Miller, G. L., \textbf{Riemann's hypothesis and tests for
primality}, J. Comput. System Sci., 13, (1976), pp 300 - 317.

\bibitem{Shor1}Shor, Peter W., \textbf{Polynomial time algorithms for prime
factorization and discrete logarithms on a quantum computer}, SIAM\ J. on
Computing, 26(5) (1997), pp 1484 - 1509. (quant-ph/9508027)

\bibitem{Shor2}Shor, Peter W., \textbf{Introduction to quantum algorithms},
\textbf{Lecture Notes for the AMS Short Course on Quantum Computation,
Washington, DC, January 2000},'' to appear in ``\textbf{Quantum Computation}%
,'' edited by S.J. Lomonaco, AMS\ PSAPM Series. (To appear) (quant-ph/0005003)

\bibitem{Stinson1}Stinson, Douglas R., ``\textbf{Cryptography: Theory and
Practice},'' CRC Press, Boca Raton, (1995). \ 
\end{thebibliography}
\end{document}